\documentclass[twocolumn,english,aps,pra,nofootinbib]{revtex4-2}

\usepackage{mathrsfs}
\usepackage{amsbsy}
\usepackage{amstext}
\usepackage[T1]{fontenc}
\usepackage[latin9]{inputenc}
\usepackage{amsmath}
\usepackage{amssymb}
\usepackage{stmaryrd}
\usepackage{graphicx}
\usepackage{color}
\makeatletter
\@ifundefined{textcolor}{}
{%
 \definecolor{BLACK}{gray}{0}
 \definecolor{WHITE}{gray}{1}
 \definecolor{RED}{rgb}{1,0,0}
 \definecolor{GREEN}{rgb}{0,1,0}
 \definecolor{BLUE}{rgb}{0,0,1}
 \definecolor{CYAN}{cmyk}{1,0,0,0}
 \definecolor{MAGENTA}{cmyk}{0,1,0,0}
 \definecolor{YELLOW}{cmyk}{0,0,1,0}
}

\usepackage{mathtools}
\usepackage{bm}
\usepackage[dvipsnames]{xcolor}

\usepackage{braket}
\renewcommand\bra[1]{{\langle{#1}|}}
\renewcommand\ket[1]{{|{#1}\rangle}}

\usepackage{dsfont}
\usepackage{natbib}
\setcitestyle{square,numbers}
\usepackage[format=plain,justification=centerlast]{caption}
\usepackage[format=plain,justification=centerlast]{subcaption}
\usepackage{textgreek}

\usepackage{babel}
\begin{document}

\title{Spectral and temporal metrology with bandlimited functions and finite-time measurements}
\author{{\L}ukasz Rudnicki}
\email{lukasz.rudnicki@ug.edu.pl}
\affiliation{International Centre for Theory of Quantum Technologies, University of Gda\'nsk, Jana Ba\.zy\'nskiego 1A, 80-309 Gda\'nsk, Poland}
\author{Tomasz Linowski}
\affiliation{International Centre for Theory of Quantum Technologies, University of Gda\'nsk, Jana Ba\.zy\'nskiego 1A, 80-309 Gda\'nsk, Poland}

\begin{abstract}
We perform an analysis supplementing the metrology toolbox in the time-frequency domain. While the relevant time-frequency-based metrological protocols can be borrowed from the spatial domain, where they have recently been well developed, their ultimate practical usefulness is shown to be restricted by limits put on the bandwidth of both the signal and measurements, as well as by the finite measurement time. As we demonstrate for the well-known problem of multiparameter estimation for two incoherent, point-like sources, the impact of these experimental limitations on the optimal protocol's efficiency can be detrimental. Nonetheless, we propose necessary operational criteria for attainability of the quantum Cram\'{e}r-Rao bound under the discussed restrictions.
\end{abstract}
\date{\today}
\maketitle

\section{Introduction}
Time and frequency form a pair of variables which is \textit{conjugate} in the sense that, already in classical signal analysis, they are connected by Fourier transform. In the quantum framework, their mutual relation is thus very similar to the one between spatial variables: position and momentum, also connected by the  Fourier transform. However, while developing a time-frequency counterpart of a notion established for the position-momentum pair, such as a toolbox for quantum metrology, careful thought due to subtle differences between the space-momentum regime and the time-frequency domain is required. 

Even though it is widely known that an operator corresponding to time does not exist (this is because frequencies are non-negative, in the same vein as energy), such a restriction does not prevent a comprehensive phase-space description of time and frequency in terms of the Wigner function \cite{Wigner_Fabre_2022}. Construction of a complete set of gates also does not pose issues, since the gates can literally be adapted from the ones in the position-momentum phase space \cite{Wigner_Fabre_2022}. 

As a consequence, the existing metrology toolbox developed for position and momentum can be also applied to problems relying on time and frequency measurements. A prime example of such a metrological scenario, which can be solved due to quantum-inspired techniques, concerns two closely separated, incoherent point-like light sources \cite{hypotheses_Lu_2018,hypothesis_testing_exoplanets_2021,hypotheses_Schlichtholz_2023}, relevant for, e.g. resolving starlight and exoplanet detection \cite{Exo_Wright_2013,Exo_Fischer_2014}. In such a case, the usual goal is to estimate the sources' relative separation \cite{superresolution_Tsang_2016,superresolution_starlight_Tsang_2019,Rayleigh_curse_Paur_2018,superrresolution_astronomy_Zanforlin_2022,crosstalk_original_PRL,crosstalk_Linowski_2023} or, in the most general, multi-parameter case, their separation simultaneously with relative brightness and the centroid \cite{unbalanced_sources_Rehacek_2017,multiparameter_Rehacek_2018}. As has been theoretically proven and experimentally verified, the very same problem phrased in time domain admits an analogous solution \cite{metrology_time_domain_Donoue_2018, time_superresolution_Ansari_2021}. Similar conclusions apply to other scenarios, such as continuous variable tomography \cite{tomography_Rehacek_2009,cv_tomography_Lvovsky_2009}: the same techniques that are used to restrict tomography protocols to a finite spatial subspace spanned by a fixed set of Hermite-Gauss (HG) or Laguerre-Gauss modes can be applied to time or frequency domain, where the HG modes are a basic tool \cite{temporal_modes_Brecht_2015}. In short, from a practical point of view, the transfer of spatial metrology techniques to time-frequency domain is mostly straightforward. 

Nonetheless, the different character of the time-frequency pair (in comparison with the position-momentum pair), does impose certain limiting factors. Taking into account the experimental conditions, in this work, we identify and analyze two salient features of metrology in the time-frequency domain:
\begin{itemize}
\item Finite bandwidth $\Omega$;
\item Finite measurement time 2$T$.
\end{itemize}
As we show, depending on the value of a single parameter $\Omega T$, the above restrictions can have significant impact on the efficiency of time-frequency protocols borrowed from the spatial domain, including the previously discussed multiparameter estimation for two incoherent sources. To derive our findings, we develop a set of tools for time-frequency metrology based on so-called \emph{spheroidal wave functions} known from signal analysis \cite{prolate_Slepian_1961,prolate_Moore_2004,prolate_review_Wang_2017}.

This paper is organized as follows. In Sec. \ref{prol}, we review the basic properties of the spherodial wave functions. In Sec. \ref{metro}, we discuss the general metrology-oriented consequences of time- and bandlimiting. These results are then applied by us in Sec. \ref{sec:limits} to derive limits on temporal superresolution obtained from well-proven protocols known from the spatial domain. Finally, in Sec. \ref{sec:summary}, we provide outlooks.

%Then, as a proof of concept, robustness of the multiparameter estimation scheme \cite{} is in Sec. \ref{} scrutinized against both aforementioned limitations. While the scheme does not anymore saturate the quantum Cramer-Rao bound, indications of its optimality (understood in a restricted sense) are provided. 

\begin{figure*}[!t]
    \centering
    \includegraphics[width=1\textwidth]{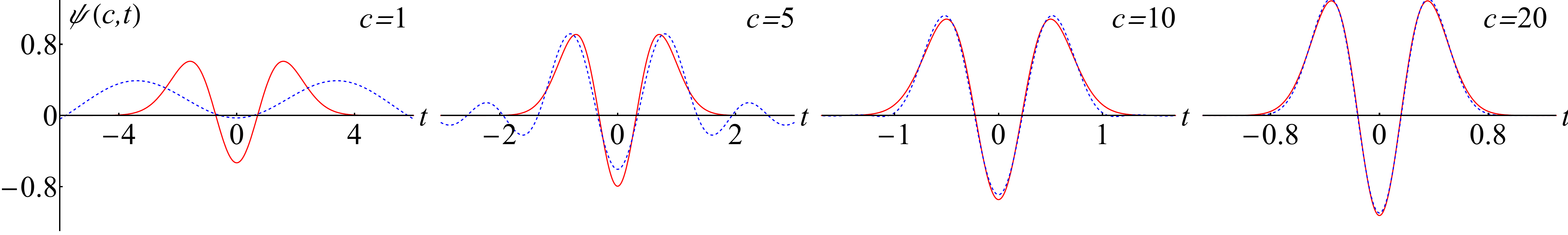}
    \caption{Comparison between the second prolate spheroidal wave function $\psi_2(c,t)$ (blue, dashed) and the second Hermite-Gauss mode $\psi_2^{\textnormal{HG}}(c,t)$ (red) for $c\in\{1,5,10,20\}$, starting from left and increasing to the right. The vertical axis is shared by all four plots. As we can see, for $c=1$ there is a severe discrepancy between the functions, while for $c=10$ the difference is almost unnoticeable.} 
    \label{Fig1}
\end{figure*}

\section{Prolate spheroidal wave functions}\label{prol}
Spheroidal wave functions of various kinds are related to the solutions of the Helmholtz equation in prolate or oblate spheroidal coordinates. Due to their unique properties, they are also ubiquitous in signal analysis, where they were introduced by Slepian \cite{prolate_Slepian_1961}. There are many sources, summarizing various useful properties of these functions. In this very brief introduction we shall rely on the pedagogical presentation from \cite{prolate_Moore_2004}, though we will slightly change the notation to fit our problem.

Let $\Omega$ be a bandwidth and $2T$ be the length of time interval
$\left[-T,T\right]$. \emph{Prolate spheroidal wave functions} (PSWF) $\psi_{n}(c,t)$
are solutions of the following integral equation:
\begin{equation}\label{Slep}
\int_{-T}^{T}dz\frac{\sin\left[\Omega\left(t-z\right)\right]}{\pi\left(t-z\right)}\psi_{n}\left(c,z\right)=\lambda_{n}\left(c\right)\psi_{n}\left(c,t\right),
\end{equation}
where $n=0,1,2\ldots$ and we stress that the domain of the PSWFs is the whole real number line, not just the interval $[-T, T]$. The dimensionless parameter $c=\Omega T$ (not to be confused with the speed of light), also sometimes called the \emph{Slepian frequency}, provides a single number characterizing the system. If $c\gg1$, potential effects due to time and band limiting are negligible. Moreover, it is enough to relax just one of these constraints to reach the limit $c\to \infty$.

The eigenvalues $0\leq\lambda_{n}\leq1$, given by the formula
\begin{equation}
    \lambda_{n}\left(c\right)=\frac{2c}{\pi}\left[R_{n0}\left(c,1\right)\right]^{2},
\end{equation}
are ordered decreasingly. In the limit $c\to \infty$, they all approach the unity, however, the largest eigenvalue $\lambda_0$ equals approximately $0.999$ already for $c=5$.

The quantities $R_{n0}(c,z)$ are radial spheroidal functions
of the first kind. While there are many platforms suitable for numerical studies of these functions, for the sake of clarity and consistency it is useful to mention that in Mathematica \cite{Mathematica}, we have the correspondence 
\[
R_{n0}\left(c,z\right)\equiv\textrm{SpheroidalS1}\left[n,0,c,z\right].
\]

The wave functions $\psi_{n}\left(c,z\right)$ are appropriately normalized
versions of $S_{n0}\left(c,z/T\right)$, with the latter being angular
spheroidal functions of the first kind. For completeness, let us mention that in Mathematica they are denoted by
\[
S_{n0}\left(c,z\right)\equiv\textrm{SpheroidalPS}\left[n,0,c,z\right],
\]
provided that the normalization of the angular functions is chosen to be
\[
\int_{-1}^{1}dz\left[S_{n0}\left(c,z\right)\right]^{2}=\frac{2}{2n+1}.
\]
Here, we adopt the following normalization of the solutions of (\ref{Slep})
\[
\psi_{n}\left(c,z\right)=\sqrt{\frac{\lambda_{n}\left(c\right)}{T\int_{-1}^{1}dz'\left[S_{n0}\left(c,z'\right)\right]^{2}}}S_{n0}\left(c,\frac{z}{T}\right),
\]
so that the relations of ``double orthogonality'' hold:
\begin{align}
\int_{-\infty}^{\infty}dz\,\psi_{n}\left(c,z\right)\psi_{m}\left(c,z\right)
    &=\delta_{nm}, \label{Orto2} \\ 
\int_{-T}^{T}dz\,\psi_{n}\left(c,z\right)\psi_{m}\left(c,z\right)
    &=\lambda_{n}(c)\,\delta_{nm}.\label{Orto1}
\end{align}
From the latter relation we can see that the eigenvalues $\lambda_{n}\left(c\right)$ quantify the amount of the corresponding PSWF contained within the interval $[-T,T]$.

An important property of PSWFs concerns their usage as an orthonormal basis for bandlimited functions. A function $f\left(t\right)$ can be expanded as
\begin{equation}\label{decomposf}
    f\left(t\right)=\sum_{n=0}^{\infty}f_n\psi_{n}\left(c,t\right)
\end{equation}
with some coefficients $f_n$ if and only if its Fourier transform $\tilde{f}\left(\omega\right)$ is supported on $\left[-\Omega,\Omega\right]$. This means that Eq. (\ref{decomposf}) is a natural way to impose the constraint of limited bandwidth on a function, which we will make frequent use of below. We remark that the expansion is valid for every finite $T$, though the coefficients $f_n$ do depend on the choice of the time limit.

%\begin{figure}[!t]
%    \centering
%    \begin{subfigure}[b]{0.23\textwidth}
%         \centering
%         \includegraphics[width=\textwidth]{c01.pdf}
%     \end{subfigure} 
%    \begin{subfigure}[b]{0.23\textwidth}
%         \centering
%         \includegraphics[width=\textwidth]{c05.pdf}
%     \end{subfigure} 
%     \hfill\\
%     \begin{subfigure}[b]{0.23\textwidth}
%         \centering
%         \includegraphics[width=\textwidth]{c20.pdf}
%     \end{subfigure}
%     \begin{subfigure}[b]{0.23\textwidth}
%         \centering
%         \includegraphics[width=\textwidth]{c10.pdf}
%     \end{subfigure}
%        \caption{Comparison between the second prolate spheroidal wave function $\psi_2(c,t)$ (blue, dashed) and the second Hermite-Gauss mode $\psi_2^{\textnormal{HG}}(c,t)$ (red) for $c\in\{1,5,10,20\}$, starting from top left and increasing clockwise. As we can see, for $c=1$ there is a severe discrepancy between both functions, while for $c=10$ the difference is almost unnoticeable.}
%        \label{Fig1}
%\end{figure}

To gain some intuition behind the above decomposition, let us point out the asymptotic connection between the PSWF and the Hermite-Gauss (HG) modes, the latter explicitly defined as
\begin{equation}
\psi_{n}^{\textnormal{HG}}\left(c,z\right)=\left(\frac{c}{\pi}\right)^{1/4}\frac{e^{-cz^{2}/2}}{\sqrt{2^{n}n!}}H_{n}\left(\sqrt{c}z\right),
\end{equation}
where the variance of the modes equals $\sigma^2=1/(2c)$ and $H_n(x)$ is the $n$-th Hermite polynomial. Then, for $c\gg 1$, it is known that \cite{prolate_asymptotic_Cloizeaux_2003}
\begin{equation}\label{eq:asym}
    \psi_{n}\left(c,z\right)\approx \psi_{n}^{\textnormal{HG}}\left(c,z\right).
\end{equation}
See Fig. \ref{Fig1} for an illustration of this effect. This result is a special case of a more general branch-dependent asymptotic expansion of the PSWFs. As we can see, if $c$ is sufficiently large, Eq. (\ref{decomposf}) is simply a standard decomposition of the function $f(t)$ in the basis formed by narrow HG modes.

\section{Metrology with time and band limitations}\label{metro}
A generic protocol of quantum metrology consists of a probe state $\hat{\varrho}_{\bm{\theta}}$, in which parameters of interest [denoted by $\bm{\theta}=(\theta_1,\theta_2,\ldots)$] are being encoded, and a positive operator valued measure (POVM), which describes the measurement used to reveal the values of the parameters \cite{metrology_noise_Zhou_2023}. The latter, for the sake of manageability, can be considered as consisting of a finite number $d$ of operators $\hat{\Pi}_i$, for $i=0,\ldots,d-1$. Due to information completeness, the last operator reads $\hat{\Pi}_d=\hat{\mathds{1}}-\sum_{i=0}^{d-1} \hat{\Pi}_i$.

Both the probe state $\hat{\varrho}_{\bm{\theta}}$ and the POVM elements $\hat{\Pi}_i$ can be decomposed in various ways in accord with the Schr{\"o}dinger mixture theorem, in particular, in terms of their eigendecompositions. Let us therefore write:
\begin{align}
    \hat{\varrho}_{\bm{\theta}}&=\sum_{k}\varrho_{k}\left|\Psi_{k}\right\rangle \left\langle \Psi_{k}\right|,\\
    \hat{\Pi}_{i}&=\sum_{l}\Pi_{il}\left|\pi_{il}\right\rangle \left\langle \pi_{il}\right|.
\end{align}
Information about the parameters $\bm{\theta}$ can be then accessed through probabilities
\begin{equation} \label{eq:p_i}
    p_{i}=\textrm{Tr}\left(\hat{\varrho}_{\bm{\theta}}\hat{\Pi}_{i}\right)
    =\sum_{k,l}\varrho_{k}\Pi_{il}\left|\left\langle \pi_{il}\left|\Psi_{k}\right\rangle \right.\right|^{2},
\end{equation}
for $i=0,\ldots,d-1$. Obviously $p_d=1-\sum_{i=0}^{d-1} p_i$. We remark that in optical considerations (like in the next section), the probe states $\hat{\varrho}_{\bm{\theta}}$ are typically restricted to mixtures of single-photon states, with $\ket{\Psi_k}$ referring to single-photon states of the associated temporal modes.

The efficiency of the protocol is quantified by the generalization of the Cram\'{e}r-Rao bound \cite{CR_bound_Cramer_1945,CR_bound_Rao_1945,noon_advances} to multiparameter estimation \cite{Cramer_Rao_bound_multiparameter_Braunstein_1994,Cramer_Rao_bound_multiparameter_Szczykulska_2016}. For a given measurement scheme, each of the parameters can be in principle measured with uncertainty not smaller than
\begin{align} \label{eq:cramer_rao}
    \Delta \theta_n \geq \sqrt{(F^{-1})_{nn}(\bm{\theta})},
\end{align}
where $F$ is the (classical) Fisher information matrix \cite{Fisher_information_original_1922}:
\begin{align} 
    F_{nm} = \sum_{j} \frac{1}{p_j}
        \frac{\partial p_j}{\partial \theta_n} \frac{\partial p_j}{\partial \theta_m}.
\end{align}
In the case of a single parameter, one can optimize over all POVMs to obtain the quantum Fisher information matrix, which saturates the Cram\'{e}r-Rao bound, corresponding to the ultimate resolution allowed by quantum mechanics \cite{Cramer_Rao_bound_multiparameter_Braunstein_1994,Cramer_Rao_bound_multiparameter_Szczykulska_2016}. In the case of more than one parameter, optimal measurements for different parameters may be mutually incompatible, preventing the Cram\'{e}r-Rao bound from being saturated \cite{QFI_attainability_Liu_2020}. Finding the general necessary and sufficient conditions under which the multiparameter quantum Fisher information matrix is attainable remains an open problem in quantum information theory \cite{five_open_problems_Horodecki_2022}. 

To incorporate both practical limitations under consideration, let us first note that in continuous variable systems, all pure states $\left|\phi\right\rangle$ can be further described in terms of wave functions. For example, in the position domain, the wave function is given by $\phi\left(x\right)=\left\langle x\left|\phi\right\rangle \right.$. The same construction does hold in time or frequency domain \cite{temporal_modes_Brecht_2015}. Without loss of generality, let us focus on the time domain. 

Following the above notation, the probe state is described by a collection of wave functions $\Psi_k(t)$, while the POVM elements are given by $\pi_{il}(t)$. Their properties are the same as for the position domain, with one important caveat: limited bandwidth implies that all functions $\Psi_k(t)$ and $\pi_{il}(t)$ are bandlimited too, i.e. their Fourier transforms ``living'' in the frequency domain must be finitely supported. This assertion holds independently of the assumed convex decompositions of the probe state and the POVMs. 

The restriction to bandlimited functions can be explicitly imposed on the wave functions associated with the probe and with the POVMs by decomposing them according to Eq. (\ref{decomposf}):
\begin{equation} \label{eq:psi_bandlimited}
    \Psi_{k}\left(t\right)=\sum_{n=0}^{\infty}\Psi_{kn}\psi_{n}\left(c,t\right),
\end{equation}
\begin{equation} \label{eq:xi_bandlimited}
    \pi_{il}\left(t\right)=\sum_{n=0}^{\infty}\pi_{iln}\psi_{n}\left(c,t\right).
\end{equation}
Then, due to the orthogonality of the PSWFs on the whole real line (\ref{Orto2}), the probabilities (\ref{eq:p_i}) take the form 
\begin{equation} \label{eq:p_i_explicit}
    p_{i}=\sum_{k,l}\varrho_{k}\Pi_{il}
        \left|\sum_{n=0}^{\infty}\pi_{iln}^{*}\Psi_{kn}\right|^{2}.
\end{equation}
Note that for large $c$ this expansion is the same as if we used narrow HG modes.

In addition to finite bandwidth, we need to account for the fact that the measurement occurs in a finite time interval, here denoted as $\left[-T,T\right]$. To this end, we introduce a projector onto this interval, denoted by $\hat{\Xi}_T$. For an arbitrary pure state $\left|\phi\right\rangle$ with its wave function $\phi\left(t\right)$, the action of this operator gives the state $\hat{\Xi}_T\left|\phi\right\rangle$, whose wave function equals $\phi\left(t\right)$ for $t\in\left[-T,T\right]$ and vanishes outside of this interval (of course, for $T\to\infty$, the wave function is untouched). Such a wave function is no longer normalized, and its norm is in general smaller than $1$. When calculating expectation values, we can think of $\hat{\Xi}_T$ as being represented by
\begin{equation} \label{eq:Xi_explicit}
    \hat{\Xi}_T = \int_{-T}^T dt \ket{t}\bra{t},
\end{equation}
with $\ket{t}$ being orthonormal vectors such that for any $\ket{\phi}$, we have $\braket{t|\phi}=\phi\left(t\right)$.

While the normalization of the probe state must be intact, the same does not need to hold for the POVM elements. At the end, the ``leakage'' element $\hat{\Pi}_d$ subsumes all information which has not been captured by the measurements. In other words, while we are not allowed to model the finite time of the measurement by truncating the probe state, we can replace the POVM elements (except the last one) according to the rule 
\begin{equation} \label{eq:POVMs_time-limiting}
    \hat{\Pi}_i \mapsto \hat{\Xi}_T \hat{\Pi}_i \hat{\Xi}_T.
\end{equation}
To incorporate this replacement inside the measurement probabilities we resort to the second orthogonality condition (\ref{Orto1}). As a result, taking into account both limitations, the probabilities (\ref{eq:p_i_explicit}) become
\begin{equation}\label{proba1}
p_{i}=\sum_{k,l}\varrho_{k}\Pi_{il}\left|\sum_{n=0}^{\infty}\pi_{iln}^{*}\Psi_{kn}\lambda_{n}\left(c\right)\right|^{2},
\end{equation}
where we stress that the equation applies only to $i=0,\ldots,d-1$. Note also the presence of $\lambda_{n}\left(c\right)$ inside the absolute value.

Physical intuition behind the steps taken above is the following. First, through Eqs. (\ref{eq:psi_bandlimited}-\ref{eq:xi_bandlimited}), we impose the restriction of being bandlimited to both the probe state and the measurements. Without this assumption, we are in principle allowed to generate arbitrarily narrowly localized signals (probe states) and perform arbitrarily narrow measurements. The second step, captured by Eq. (\ref{eq:POVMs_time-limiting}), accounts for the fact that the measurement does not last ``forever''. So even though the signal can extend beyond (especially before) the measurement starts, the measurement itself happens
in a finite time window. 

As previously discussed, the extent to which the two restrictions affect the measurement is ultimately captured by a single parameter, the Slepian frequency $c$. In \cite{time_superresolution_Ansari_2021}, for example, the bandwidth and measurement time characterizing the experiment equaled 17 GHz and 20 {\textmu}s, resulting in $c\approx 10^5$, way above the negligibility threshold $c\gg 1$. However, in that same paper, the true accessible time scale was reported to be of the order of 30 fs, which would correspond to the radically small $c\approx 10^{-6}$, given the same bandwidth. With a growing need for enhanced measurement efficiency and precision, we predict the latter regime to be gradually approached, necessitating taking the discussed restrictions into account.

\begin{figure}[!t]
    \centering
    \includegraphics[width=0.49\textwidth]{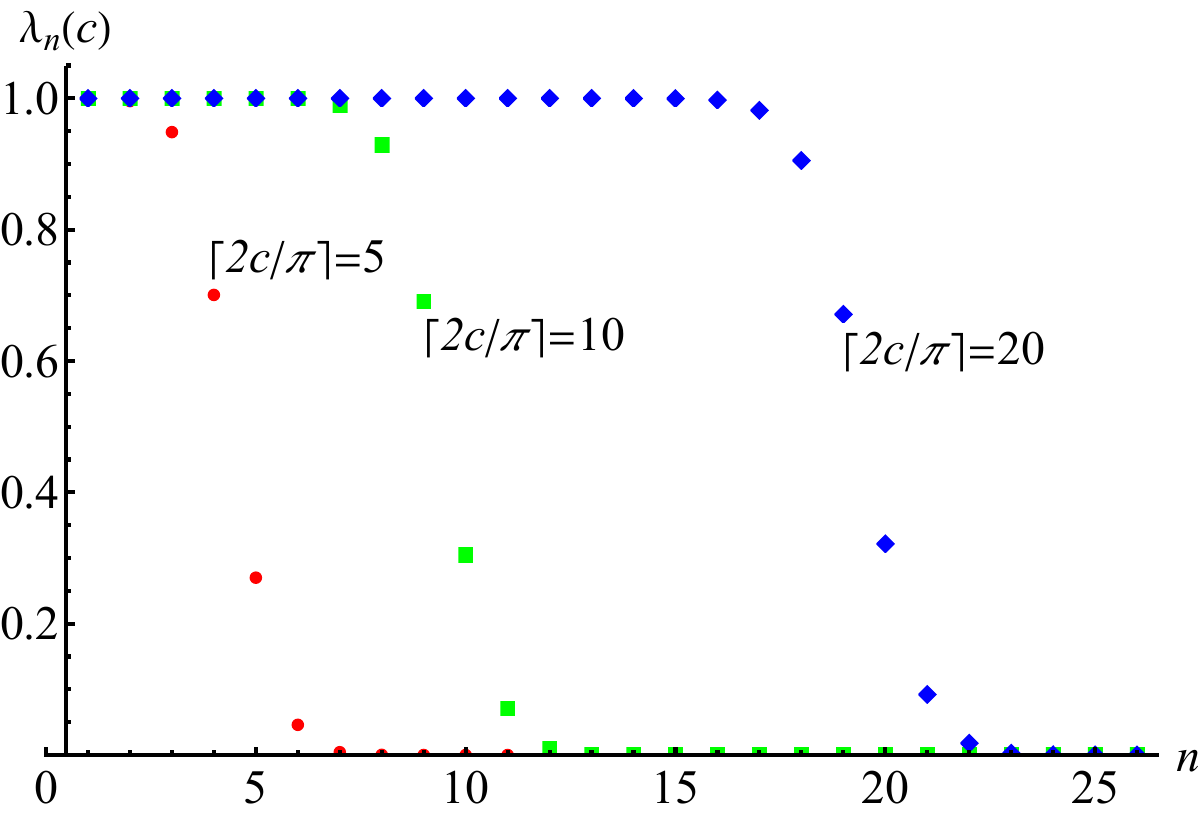}
    \caption{Eigenvalues $\lambda_{n}\left(c\right)$ shown as a function of $n$ (on the horizontal axis), for $c$ such that $\left\lceil \left(2c\right)/\pi\right\rceil=\{5,10,20\}$ (red circles, green squares, blue diamonds, respectively). As can be seen, the eigenvalue is almost equal to unity starting from $n=0$ and up to approximately $n=\left\lceil \left(2c\right)/\pi\right\rceil$, around which we observe a sudden decay involving three to four eigenvalues. The rest of the spectrum is approximately vanishing.} 
    \label{Fig2}
\end{figure} 

To infer meaningful conclusions of general validity we resort to a very handy property of the eigenvalues $\lambda_{n}\left(c\right)$. As is well known \cite{prolate_Moore_2004}, with the exception of relatively small values of $c$, the first $\left\lceil \left(2c\right)/\pi\right\rceil$ eigenvalues are very close to $1$, while the remaining eigenvalues quickly drop to zero. This behaviour can be observed in Fig. \ref{Fig2}. Given this, our previous result (\ref{proba1}) can be faithfully approximated as
\begin{equation}\label{proba2}
p_{i}\approx\sum_{k,l}\varrho_{k}\Pi_{il}\left|\sum_{n=0}^{\left\lceil \left(2c\right)/\pi\right\rceil}\pi_{iln}^{*}\Psi_{kn}\right|^{2}.
\end{equation}
An underpinning of this equation can serve as a litmus paper for optimality in a quantum metrology protocol, the latter being understood as saturation of the quantum Cramer-Rao bound (\ref{eq:cramer_rao}). 

To this end, given a probe state, one shall look for a POVM saturating this bound. Assuming that such a POVM has been found, we need to expand it in the PSWF basis. As long as it is enough to consider only first $\left\lceil \left(2c\right)/\pi\right\rceil$ terms of this expansion, optimality is preserved.  In other words, we shall require that the optimal POVM fulfills
\begin{equation}
    \pi_{il}\left(t\right)\approx
    \sum_{n=0}^{\left\lceil \left(2c\right)/\pi\right\rceil}\pi_{iln}\psi_{n}\left(c,t\right).
\end{equation}
This is true because probabilities relevant for the discussed measurement circumstances, which turn out to be  well approximated by (\ref{proba2}), do not differ from the probabilities obtained in an infinitely long experiment. However, when higher order terms are necessary to describe the optimal POVM elements faithfully, we know that optimality is lost. What might be even more striking, is that there is no other choice for a POVM, which would be able to restore optimality. Had that been possible, we would have already started with an optimal POVM sufficiently well approximated by the terms of the order not higher than $\left\lceil \left(2c\right)/\pi\right\rceil$.

\section{Limits on temporal superresolution} \label{sec:limits}
To demonstrate the potential effect of time- and bandlimiting in practical scenarios, we consider the problem of resolving temporal separations between two point-like incoherent light pulses.

\begin{figure}[!t]
\centering
\includegraphics[width=0.49\textwidth]{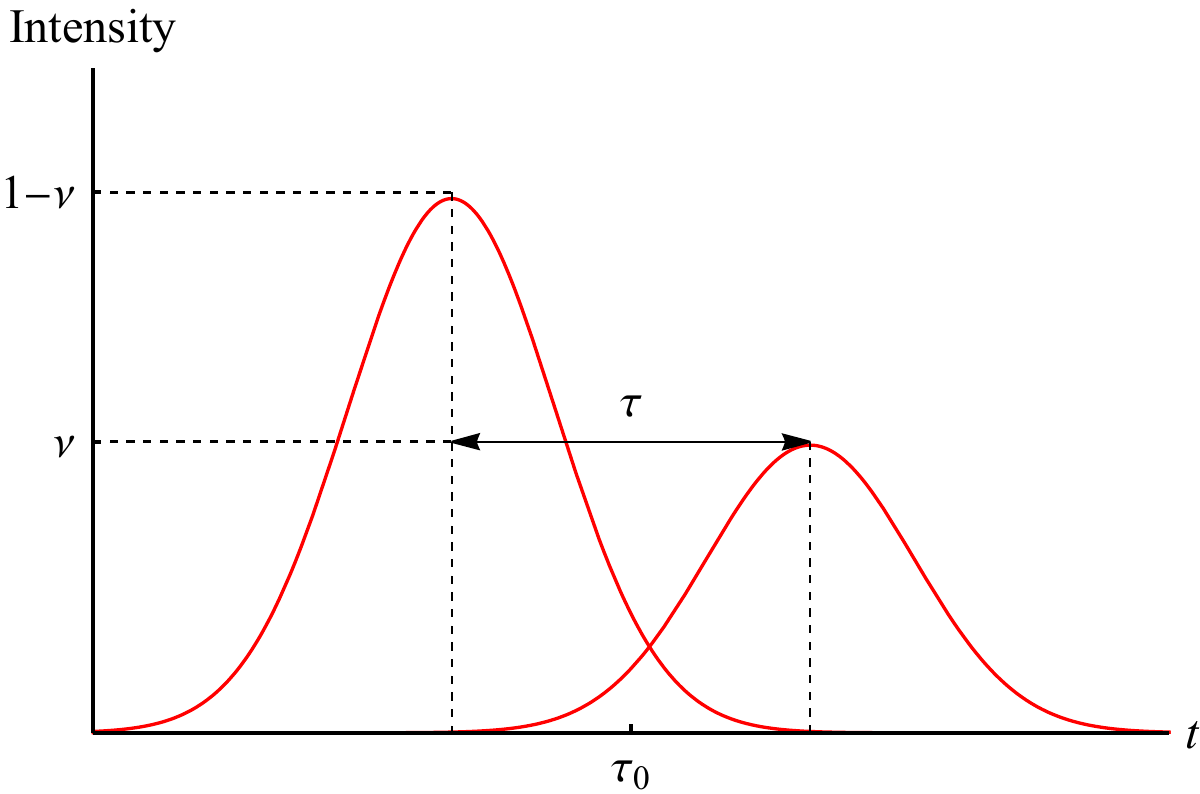}
\caption{Illustration of the problem of resolving temporal separations between two point-like incoherent light pulses. Two incoherent signals characterized by different intensities $\nu$ and $1-\nu$ are centered at $\tau_0$ and temporally separated by $\tau$ (the latter assumed to be very small). The goal is to simultaneously estimate all three parameters with the best possible precision. In the optimal scheme, the measurement consists of projections onto a set constructed from the Hermite-Gaussian modes, like the one in Fig. \ref{Fig1}.} 
\label{fig:illustration}
\end{figure}

As discussed previously in the Introduction, in the absence of time and band limitations, the scenario is fully analogous to the well-known problem of resolving spatial separations between light sources \cite{superresolution_Tsang_2016,superresolution_starlight_Tsang_2019,Rayleigh_curse_Paur_2018,superrresolution_astronomy_Zanforlin_2022,crosstalk_original_PRL,crosstalk_Linowski_2023,unbalanced_sources_Rehacek_2017,multiparameter_Rehacek_2018}, and hence can be modeled in a similar fashion \cite{metrology_time_domain_Donoue_2018, time_superresolution_Ansari_2021}. Assuming that the pulses have relative intensities $\nu$ and $1-\nu$ (so that the total intensity is normalized to one) and that they are temporally separated by $\tau$ with centroid $\tau_0$, the total signal is represented by the following density operator \cite{multiparameter_Rehacek_2018}:
\begin{align} \label{eq:rho}
\begin{split}
    \hat{\varrho}_{\bm{\theta}} 
        = \nu\ket{\Psi_+}\bra{\Psi_+} + (1-\nu)\ket{\Psi_-}\bra{\Psi_-},
\end{split}
\end{align}
with the parameters of interest being $\bm{\theta}=(\tau,\tau_0,\nu)$. Here, $\ket{\Psi_\pm}$ are $\tau$-displaced states of the amplitude point spread function $\Psi(t)$ describing the pulses, defined in such a way that $\braket{t|\Psi_\pm}=\Psi_\pm(t)=\Psi(t-\tau_0\mp\tau/2)$, with $\tau_0$ being the signal's centroid. See Fig. \ref{fig:illustration} for illustration. Note that $\Psi(t)$ is assumed to be real-valued. We remark that, although usually only the separation is of practical significance, in practice it cannot be reliably measured without simultaneously treating the remaining two parameters, hence the need for multiparameter estimation. 

For the problem at hand, it was found in \cite{multiparameter_Rehacek_2018} that the quantum Fisher information, saturating the Cram\'{e}r-Rao bound (\ref{eq:cramer_rao}), is achieved with the following scheme. To start with, one needs to define a basis in the signal space. A convenient choice arises by considering the set $\{\ket{\Gamma_n}\}$, defined by
\begin{align}
    \braket{t|\Gamma_n} = \frac{\partial^n}{\partial t^n}\Psi(t-\tau_0).
\end{align}
From this, an orthonormal basis $\{\ket{\Phi_n}\}$ is obtained by means of the standard Gram-Schmidt process.
Then, the optimal POVM set consists of four elements: the first three ($j=0,1,2$) being projectors $\hat{\Pi}_j=\ket{\pi_j}\bra{\pi_j}$ of the form 
\begin{align}
    \ket{\pi_j}\equiv\sum_{k=0}^3 C_{jk}\ket{\Phi_k},
\end{align}
and the last defined as $\hat{\Pi}_3 = \hat{\mathds{1}}-\hat{\Pi}_0-\hat{\Pi}_1-\hat{\Pi}_2$. Here, $C_{jk} = \braket{\Phi_k|\pi_j}$ are real coefficients, such that the POVMs are all non-negative and linearly independent.

As was found in \cite{multiparameter_Rehacek_2018}, provided a suitable choice of alignment for the measurement apparatus and as long as\footnote{According to Eq. (13) in \cite{multiparameter_Rehacek_2018}, the non-vanishing coefficients should be $C_{20},C_{21}$, rather than $C_{02},C_{12}$. However, it is clear from their further results that the latter was meant.}
\begin{align} \label{eq:C_conditions}
C_{00}=C_{10}=0, \quad C_{01},C_{11},C_{02},C_{12}\neq 0,
\end{align} 
the Fisher information obtained from this scheme differs from the quantum Fisher information only by a multiplicative factor of
\begin{align} \label{eq:A}
    \mathscr{A}\equiv\frac{(C_{01}C_{12}-C_{11}C_{02})^2}{C_{01}^2+C_{11}^2}\leq 1.
\end{align}
Being dependent on four parameters, one may suspect that the inequality can be easily saturated. Indeed, as reported in \cite{multiparameter_Rehacek_2018}, this can be done in infinitely many ways, meaning that there are infinitely many measurements attaining the quantum limit. As we will now show, for measurement that is time- and bandlimited, this is no longer possible.

\begin{figure}[!t]
\centering
\includegraphics[width=0.49\textwidth]{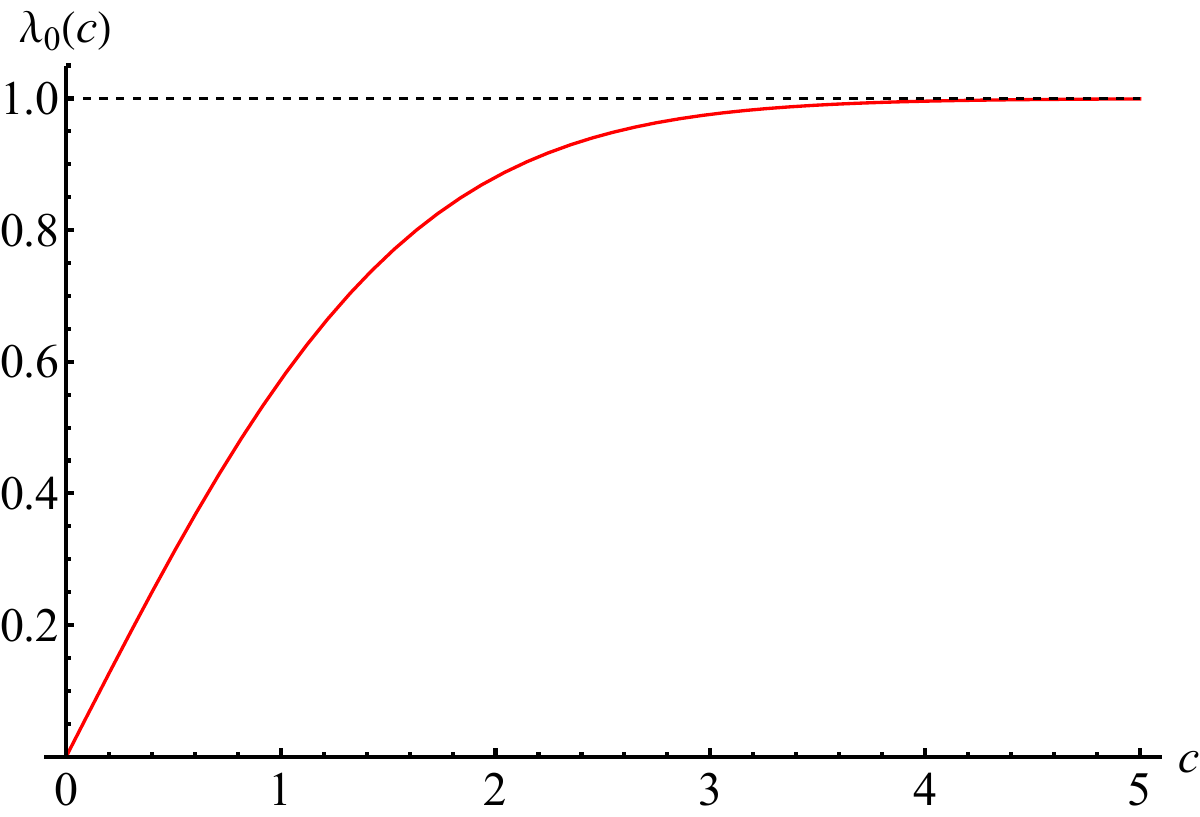}
\caption{The largest eigenvalue of PSWFs $\lambda_0(c)$ (red), constituting an upper bound on the multiplicative factor by which the Fisher information in the optimal measurement scheme differs from the ultimate quantum precision in the presence of band- and time-limiting. For convenience, the ideal value of the multiplicative factor, equal to one, is denoted by a dashed black line. As seen, for values of $c$ close to zero, the limitations put on superresolution are severe. \label{fig:lambda0}}
\end{figure}

To see this, let us first recall how the above limitations affect the measurement. Firstly, due to bandlimiting, all the time-dependent functions have to be decomposable into PSWFs, as in (\ref{decomposf}). Notably,
\begin{align} \label{eq:Phi_bandlimited}
     \Phi_{k}(t) = \sum_{n=0}^{\infty}\Phi_{kn}\psi_{n}\left(c,t\right).
\end{align}
Secondly, due to time-limiting, the POVMs should be replaced according to the rule (\ref{eq:POVMs_time-limiting}). In particular, this implies the following change in the coefficients defining the POVMs:
\begin{align}
     C_{jk} = \braket{\Phi_k|\pi_j} \to \bra{\Phi_k}\hat{\Xi}_T\ket{\pi_j}.
\end{align}
This is all we need to prove our claim. Because all the coefficients are real, we have
\begin{align}
     C_{jk}^2 = C_{jk} C_{jk}^* 
        = \bra{\Phi_k}\hat{\Xi}_T\hat{\Pi}_j\hat{\Xi}_T\ket{\Phi_k}.
\end{align}
In particular,
\begin{align}
     C_{01}^2 + C_{11}^2 = \bra{\Phi_1}\hat{\Xi}_T(\hat{\Pi}_0+\hat{\Pi}_1)\hat{\Xi}_T\ket{\Phi_1} 
        \leq \bra{\Phi_1}\hat{\Xi}_T\ket{\Phi_1},
\end{align}
where the last step follows from the fact that $\hat{\Pi}_0+\hat{\Pi}_1\leq \hat{\mathds{1}}$. The above inequality defines a two-dimensional sphere in the space of $(C_{01},C_{11})$ centered at $(0,0)$ and with radius squared $r_1^2\leq \bra{\Phi_1}\hat{\Xi}_T\ket{\Phi_1}$. This means that we can parametrize $C_{01}$, $C_{11}$ as
\begin{align}
     C_{01} = r_1 \sin\phi_1, \quad C_{11} = r_1 \cos\phi_1,
\end{align}
where $\phi_1\neq n\pi/2$, so that neither coefficient vanishes, as required by (\ref{eq:C_conditions}). In full analogy, for some $r_2^2\leq \bra{\Phi_2}\hat{\Xi}_T\ket{\Phi_2}$ and $\phi_2\neq n\pi/2$,
\begin{align}
     C_{02} = r_2 \sin\phi_2, \quad C_{12} = r_2 \cos\phi_2.
\end{align}
Substituting these into (\ref{eq:A}) and using elementary properties of trigonometric functions, we obtain
\begin{align}
\begin{split}
    \mathscr{A} &= r_2^2 \sin^2(\phi_1-\phi_2) \leq r_2^2 \\
        &\leq \bra{\Phi_2}\hat{\Xi}_T\ket{\Phi_2} = \int_{-T}^T dt \, \Phi_2^2(t),
\end{split}
\end{align}
where in the last step we used (\ref{eq:Xi_explicit}). According to the Schwartz's Paley-Wiener theorem \cite{Schwartz_theorem_1952}, any function (other than the constant zero function), whose Fourier transform has finite support, cannot be finitely-supported itself. This means that, since $\Phi_2$ is bandlimited, the r.h.s. above is strictly smaller than unity for any finite $T$, finishing the proof.

A more non-trivial upper bound follows from explicit calculation. Using (\ref{eq:Phi_bandlimited}) and (\ref{Orto1}) in the last equation, we get
\begin{align}
\begin{split}
    \mathscr{A} \leq \sum_{n=0}^\infty \Phi_{2n}^2 \lambda_n(c).
\end{split}
\end{align}
Since $\lambda_n(c)$ are ordered decreasingly, we can bound the sum from above by replacing all $\lambda_n(c)$ with $\lambda_0(c)$, yielding
\begin{align} \label{e:A_final_inequality}
\begin{split}
    \mathscr{A} \leq \lambda_0(c).
\end{split}
\end{align}
Therefore, in the presence of band- and time-limiting, the Fisher information attained from the optimal scheme differs from the ultimate precision allowed by quantum mechanics by a multiplicative factor not larger than the eigenvalue $\lambda_0(c)$. As seen from Fig. \ref{fig:lambda0}, despite approaching unity relatively quickly with $c$, there is still a significant range of values of this parameter, for which the implied limitations are severe.

Let us furthermore observe that the inequality (\ref{e:A_final_inequality}) is saturated only if $\Phi_{2m}=\delta_{m,0}$, i.e. if $\Phi_{2}(t)$ coincides with the zeroeth PSWF $\psi_0(c,t)$. However, this is highly unlikely as, by construction, $\Phi_{2}(t)$ is related to the second derivative of the amplitude point spread function, which should be closer to the second or higher PSWF. For example, in the well-known case of a Gaussian point spread function, the ideal $\Phi_{2}(t)$ corresponds to the second Hermite-Gauss mode, which, due to the asymptotic relation (\ref{eq:asym}), must reduce to a function much closer to the second than the zeroeth PSWF after bandlimiting. In short, we expect that in practice, the true implications of time- and bandlimiting should be even more severe than demonstrated here. As such, in experiment, special attention should be paid to extend both the bandwidth and the time of each measurement, so as to maximize the efficiency-lowering parameter $c$.

\section{Outlooks} \label{sec:summary}
Virtually all techniques developed for metrology in the spatial domain can be recontextualized to the time (or frequency) domain in a way which, at least from the theory perspective, is straightforward. However, our analysis, taking into account potential experimental limitations, reveals that during such adaptation, the optimality of a given metrology protocol is lost if the dimensionless product of the bandwidth and the time of the measurement $c$ (the Slepian frequency) is too small in comparison with the number of prolate spheroidal wave functions (PSWFs) with the same Slepian frequency necessary to faithfully reproduce the optimal POVMs. This was demonstrated by us explicitly for multiparameter estimation for two incoherent, point-like source, where small values of $c$ had radical impact on protocol efficiency.

While typical contemporary experiments should be characterized by sufficiently large Slepian frequency, practical problems, such as the need to maximize the number of measurements performed during an allocated time window, or simply lack of appropriate resources, may lead to extreme experimental conditions, in which both the bandwidth and time of the measurement are pushed to its limits. It is clear that in such a regime, strategies different to the ones from the spatial domain need to come into play. From a comparison between (\ref{proba1}) and (\ref{proba2}), we can see that a partial solution to the problem is given by a careful preparation of the probe state. If the latter is approximately decomposable into exclusively low order PSWFs, the optimal POVMs will follow this pattern, leading to high protocol efficiency. Specific methods to counteract the detrimental effect of band- and time-limiting shall be considered as the right target for further studies.

\section*{Acknowledgments}
Discussions with L. L. S{\'a}nchez-Soto, Z. Hradil and J. \u{R}eh{\'a}\u{c}ek are gratefully acknowledged. This project has received funding from the European Union's Horizon 2020 research and innovation programme under grant agreement No 899587 (STORMYTUNE).

\bibliography{main}
\bibliographystyle{obib}

\end{document}